\documentstyle[tighten,aps,emlines,eqsecnum]{revtex}
\begin{document}
\author{V.V.Gafiychuk}
\address{Instute for Applied Problems of Mechanics and Mathematics,\\
National\\Academy of Sciencies of Ukraine, 3b~Naukova~str.,
Lviv,\\290601,\\Ukraine}
\author{I.A.Lubashevsky}
\address{Laboratory of Synergetics, the Moscow State University, Vavilova\\str.\\
46/92, Moscow, 117333, Russia}
\author{Robert E. Ulanowicz}
\address{University of Maryland Chesapeake Biological\\ Laboratory,
Solomons,\\MD\\20688-0038 USA}
\title{Distributed Self-regulation Induced by Negative Feedbacks in Ecological and
Economic Systems}
\date{\today}
\maketitle

\begin{abstract}
We consider an ecological system governed by Lotka-Volterra dynamics and an
example of an economic system as a mesomarket with perfect competition. We
propose a mechanism for cooperative self-regulation that enables the system
under consideration to respond properly to changes in the environment. This
mechanism is based on (1) active individual behavior of the system elements
at each hierarchical level and (2) self-processing of information caused by
the hierarchical organization. It is shown how the proposed mechanism
suppresses nonlocal interaction of elements belonging to a particular level
as mediated by higher levels.
\end{abstract}

%\author{Vasyl V. Gafiychuk, Ihor A. Lubashevsky, and Robert E. Ulanowicz}

\section{Introduction}

A great number of natural systems are organized hierarchically. Their
hierarchical organization allows that such a system can be divided into a
collection of subsystems (which will be called levels) involving many
elements that are similar in their properties. The elements of the various
levels differ substantially, however, in their characteristics. The
subsystems can be ordered according to their mutual interactions: The
behavior of an element at each level is determined by the aggregated state
of a certain large group of elements belonging to the nearest lower level,
while each element of a lower level is directly governed by a given element
of the higher level.

Such hierarchical organization is inherent in many ecological and economic
systems. For example, we encounter a huge number of goods in an economic
market in contrast to relatively few types of raw materials. Hence, the
network of products and trade that transforms natural materials into a wide
variety of goods will be a highly branching system. Suppose firms of a given
type of activity that are approximately equal in power make up a certain
level. The market then involves several such levels, from lower ones
consisting of retailing companies up to the highest one that deals with the
production of raw materials. In this case, the prices of products of firms
dealing in wholesale trade are the direct averages of the prices of goods at
the terminal retail points that are supplied by these firms.

Hierarchical organization is encountered frequently in ecological systems as
well. Ecological systems often form trophic food chains or pyramids. Levels
of such an ecosystem are made up of animals comparable in size and playing
much the same role in the prey-predator relationships. Energy usually flows
from smaller organisms via consumption to larger predators. The linkages
from the small organisms generally vary over smaller scales. The larger
animals that dominate these smaller organisms do so over larger scales of
space and time. That is, because of their wider ambits, predators control
larger regions of space for longer times. From this perspective, the
hierarchical levels of most pelagic trophic networks are defined according
to particle size (Platt et al.1981). It becomes possible to regard the
populations at each level as being continuously distributed across their
particular segment of space. This representation of the trophic hierarchy is
depicted in Fig.1.

The characteristic feature of hierarchical systems is the nonlocal
interference among elements at the same level as mediated by the higher
levels. Higher levels in their turn feel only the averaged state of the
preceeding levels. Thus, local variations in the behaviors of elements
belonging to lower levels reflect the states of elements at higher level
over larger scales. The larger component then changes the state of elements
at the lower level in a region whose domain substantially exceeds the size
of the initial perturbation. Such nonlocal interaction is not reliable,
because it does not stem from the local laws of element interaction between
neighboring levels that control the life of the system.

These characteristics make such hierarchical systems fragile with respect to
perturbations in the environment. In order for systems to continue living,
there should be some mechanism of self-regulation that would maintain system
stablity and would suppress (at least, to some extent) nonlocal interactions
among elements at the same level.

Below we suggest a possible mechanism for such self-regulation. In
particular, in the next section we analyze a set of model ecosystems
governed by Lotka-Volterra dynamics. Certainly, the dynamics of real
ecosystems are far more sophisticated, however, models with Lotka-Volterra
dynamics typify many ecosystem characteristics and highlight nonlocal
fragility in a most pronounced way.

A cooperative mechanism for self-regulation whereby the hierarchical system
as a whole can react perfectly has been developed by\cite{ec4},\cite{e4a}.
This mechanism consists in the response of each individual element to the
small piece of the information available to it on the state of the whole
system. The conservation of medium flowing through the supplying network
gives rise to a certain processing of information that results in
self-consistent behavior of the elements that leads to perfect
self-regulation.

\section{Ecosystem model and the distributed self-regulation}

We begin by considering a simple mathematical model of a pelagic marine
ecosystem involving 2$N$ levels in which is found a large number of animal
species. At the bottom of this system is phytoplankton (level 1) and at the
top (level 2$N$) stands the population of large predatory fish. The
characteristic features that distinguish each level, for example level $i$,
are the body size of the individual organisms and the spatial size $\ell _i$
of the domain that is controlled by each individual fish at this level.

The flow of biomass in this trophic system is assumed to be governed by the
Lotka-Volterra model, which describes hierarchical level $i$ in terms of the
spatial distribution of the biomass $c_i({\bf r},t)$ and treats the
interaction between different levels as feeding relations, where the larger
species play the role of predators and the smaller, those of prey. According
to what was discussed in the Introduction, we assume that the characteristic
lengths $\{\ell _i\}$ of the control by individuals meet the following
inequalities 
\begin{equation}
\label{1.1}\ell _1\ll \ell _2\ll ...\ll \ell _{2N} 
\end{equation}
This assumption may be justified on allometric grounds, i.e. most
physiological processes scale as an algebraic power of body size. Here we
are extending the allometric notion to include the ambits of the organisms
in question (Zotin 1985,\ Cousins 1985).

The dimensionless distribution $c_i({\bf r},t)$ is governed by the equation 
\begin{equation}
\label{1.2}\tau _i\frac{\partial c_i}{\partial t}=\left(
c_ic_{i-1}-c_ic_{i+1}+\alpha \delta _{i1}-\beta _ic_i\right) -\ell _i\nabla
J_i. 
\end{equation}
where $\tau _i$, $\beta _i$ are given constants, and the term $\alpha \delta
_{i1}$ ($\delta _{i1}$ is the Kroneker symbol) describes the input of
biomass through the first level (phytoplankton) . Equation (\ref{1.2}) is an
example of the standard form of the Lotka-Volterra dynamics as applied to a
linear trophic chain (with the exception of the last term on the right-hand
side.) The final term describes the dynamics of nonuniformities in the
spatial distribution of species $i$, where $J_i$ is the movement of its
members through space. Usually, the relationship between the $J_i$ and
nonuniformites in their distributions, $c_i({\bf r},t)$, is written in the
form (Svirezhev,1987 ) 
\begin{equation}
\label{1.3}J_i=-\ell _i\nabla c_i. 
\end{equation}
Expression (\ref{1.3}) actually corresponds to the passive behavior of
animals undergoing random motion in space and independent both of other
members of the same species and of their predators and prey. In this paper
we account for the active behavior of animals at every hierarchical level.
This means that each animal attempts, (1) to avoid any region where the
concentration of members of the same species is large, in order to decrease
the competition for feed resources, (2) to prefer to visit domains
containing high concentrations of prey and, (3) to avoid regions with many
predators. Such active behavior will be described by the following
expression: 
\begin{equation}
\label{1.4}J_i=\ell _i\left[ -(1+\omega _{i,i}c_i)\nabla c_i-\omega
_{i,i+1}c_i\nabla c_{i+1}+\omega _{i,i-1}c_i\nabla \left\langle
c_{i-1}\right\rangle _{\ell _{i-1}}\right] 
\end{equation}
where $\omega _{i,i}$, $\omega _{i,i+1}$, and $\omega _{i,i-1}$ are positive
constants and $\left\langle c_{i-1}\right\rangle _{\ell _{i-1}}$ is the
concentration of prey averaged over the domain of their individual
lifespans. Let us specify the value of $\left\langle c_i\right\rangle _{\ell
_i}$ by the expression 
\begin{equation}
\label{1.4a}\left\langle c_i\right\rangle _{\ell _i}({\bf r})=\int d{\bf r}%
^{\prime }A\exp \left\{ \frac{({\bf r}-{\bf r}^{\prime })^2}{2\pi \ell _i^2}%
\right\} c_i({\bf r}) 
\end{equation}
where $A$ is a normalization constant. The nonlinear terms in expression~(%
\ref{1.4}) are those responsible for the self-regulation. It should be noted
that a similar expression for $J_i$ has been used by V.V. Alexeev (1976) and
P.S.Landa (1983) to describe the active behavior of zooplankton.

Let us justify the assumptions on active behavior by analyzing a
steady-state small perturbation in the uniform distribution $\{c_i^0\}$ of
the given species in space under the constraints 
$$
c_i^0c_{i-1}^0-c_i^0c_{i+1}^0+\alpha \delta _{i1}-\beta _ic_i^0=0. 
$$
Linearizing equation (\ref{1.2}) and expression (\ref{1.4}) with respect to
steady-state perturbation $\delta c_i\propto \exp (i{\bf kr})$, we get for $%
i\geq 2$

\begin{equation}
\label{1.5}-k^2\ell _i^2[(1+\omega _{i,i}c_i^0)\delta c_i+\omega
_{i,i+1}c_i^0\delta c_{i+1}-F(k\ell _{i-1})\omega _{i,i-1}c_i^0\delta
c_{i-1}]+c_i^0\delta c_{i-1}-c_i^0\delta c_{i+1}=0 
\end{equation}
where $F(k\ell _i)=\exp \left\{ -\frac 12k^2\ell _i^2\right\} $ is the
Fourier transform of the kernel of integral operator (\ref{1.4a}).

Let us analyze in particular how a perturbation occurring initially at a
lower level propagates through the trophic system to its highest levels and
the opposite case, i.e., a perturbation moving from top to bottom. In the
first case it is useful to introduce the quantities

$$
f_i=\frac{\delta c_i/c_i^0}{\delta c_{i-1}/c_{i-1}^0} 
$$
that relate the relative values of perturbation at one level with those at
the nearest neigbouring levels. This allows us to rewrite expression (\ref
{1.5}) in the form

\begin{equation}
\label{1.6}k^2\ell _i^2(1+\omega _{i,i}c_i^0)+(1+k^2\ell _i^2\omega
_{i,i+1})c_{i+1}^0f_{i+1}=(1+k^2\ell _i^2F(k\ell _{i-1})\omega
_{i,i-1})c_{i-1}^0f_i^{-1}. 
\end{equation}
In order to analyze the propagation of the perturbation in the chosen
direction we may set $\delta c_{2N}=0$ \cite{Her}, that is $f_{2N}=0$. So
for $i=2N-1$

\begin{equation}
\label{1.7a}f_{2N-1}=\left. \frac{(1+k^2\ell _i^2F(k\ell _{i-1})\omega
_{i,i-1})c_{i-1}^0}{k^2\ell _i^2(1+\omega _{i,i}c_i^0)}\right| _{i=2N-1} 
\end{equation}
and for $1<$ $i<2N-1$%
\begin{equation}
\label{1.7b}f_i=\frac{(1+k^2\ell _i^2F(k\ell _{i-1})\omega _{i,i-1})c_{i-1}^0%
}{k^2\ell _i^2(1+\omega _{i,i}c_i^0)+(1+k^2\ell _i^2\omega
_{i,i+1})c_{i+1}^0f_{i+1}}. 
\end{equation}
In order to analyze the behavior of the quantities $f_i$ as the level $i$
changes, we fix the wave number $k$ such that $k\ell _{i^{*}}\ll 1$, whereas 
$k\ell _{i^{*}+1}\gg 1$ for a particular level $i^{*}$ (for example, $%
k=(\ell _{i^{*}}\ell _{i^{*}+1})^{-1/2}$). As follows from (\ref{1.7b}), for 
$i<i^{*}$ the values $f_i$ and $f_{i+1}$ are related by the expression 
$$
f_i=\frac{c_{i-1}^0}{c_{i+1}^0f_{i+1}} 
$$
and for $i>i^{*}+1$, the value $f_i\ll 1$. The magnitude of the quantity $%
f_{i^{*}+1}$ depends substantially on the parameter $\omega _{i^{*}+1,i^{*}}$%
. Indeed, if $\omega _{i^{*}+1,i^{*}}=0,\;\omega _{i^{*}-1,i^{*}}=0$, the
value $f_{i^{*}+1}\ll 1$, whereas for $\omega _{i^{*}+1,i^{*}}\sim 1$, we
get $f_{i^{*}+1}\sim 1$, too. Therefore in the first case, which corresponds
to the passive behavior of animals, the quantities $f_{i^{*}}$, $f_{i^{*}-2}$%
, $f_{i^{*}-4}$, $\ldots $ are large. This last condition means that the
relative variations of the concentrations $\delta c_2/c_2$, $\delta c_4/c_4$%
, $\ldots ,$ can be large in comparison with the perturbation $\delta
c_1/c_1 $ occuring at the bottom of the trophic system. In other words, the
passive ecosystem is fragile. When the animals exhibit active behavior,
however, all the values $f_i$ for $i<i^{*}$ are of order unity, so that a
small perturbation at the bottom of the ecosystem cannot lead to substantial
perturbations at other levels. This is the essence of the proposed mechanism
for self-regulation.

It should be noted that perturbations of lower levels lead to responses with
consistently the same signs going up the food chain toward top carnivores
(big animals).This agrees with the results obtained by Herendeen, 1996.
Indeed, stock changes in the ecosystems under consideration can be
represented as $([+],s,s,...s)$, signifying that changes in stocks of the
producer (the bracketed term) lead to stock changes at successively higher
levels that have the same sign as that of the perturbed lower compartment
(Here ''s'' means the perturbation has the same sign.)

Let us now consider the characteristics of the propagation of perturbations
from the top to the bottom. In this case it makes sense to consider only
perturbations characterized by a spatial scale comparable with the size $%
\ell _{2N}$ of the domain controlled by the largest predators, that is, we
may assume that $k\ell _i\ll 1$ for practically all the levels. Under such
conditions we may set $\delta c_1=0$ \cite{Her}, and it is useful to
introduce the quantities $\{f_i\}$ specified by the expression

$$
f_i=\frac{\delta c_i/c_i^0}{\delta c_{i+1}/c_{i+1}^0}, 
$$
which allows us to rewrite equation (\ref{1.5}) as follows 
\begin{equation}
\label{1.10}k^2\ell _i^2(1+\omega
_{i,i}c_i^0)+c_{i+1}^0f_i^{-1}=c_{i-1}^0f_{i-1}. 
\end{equation}
In a similar way we get 
$$
f_1=0, 
$$
\begin{equation}
\label{1.11}f_2=-\left. \frac{c_{i+1}^0}{k^2\ell _i^2(1+\omega _{i,i}c_i^0)}%
\right| _{j=2}, 
\end{equation}
and for $i>2,$ 
\begin{equation}
\label{1.12}f_i=\frac{c_{i+1}^0}{c_{i-1}^0f_{i-1}-k^2\ell _i^2(1+\omega
_{i,i}c_i^0)}. 
\end{equation}
Whence it follows that the changes in stocks can be represented as $%
(...o,s,o,[+]),$where an increase in the stocks of the top carnivores
alternates the sign of the perturbations going down the chain. (''0'' means
that the perturbation has the opposite sign). In addition, the values $f_i$
alternate between small and large as we pass through the levels. In other
words, ecosystems configured as trophic chains cannot effectively regulate
themselves with respect to perturbations in populations of the large
predators. This difficulty does not pertain, however, to our postulated
mechanism of self-regulation, which suppresses nonlocal interaction of lower
level elements as mediated by the higher levels. In general, our results
accord with the consensus among ecologists that bottom-up control tends to
be stabilizing, whereas top-down influences are usually destabilizing.

\section{Self-regulation in a market with perfect competition}

In this section we create a simple, distributed model of a market in which
the price of each type of goods does not depend on the demand for goods of
other types. In other words, in such a market there is no nonlocal
interaction of the flows of different goods, which is due to the mechanism
of self-regulation to be considered. In this context it is reasonable to
confine ourselves to a mesomarket of goods made primarily from the same raw
material. Hence, this market will involve a single network that joins the
ultimate consumers with all types of producers, including the firms
producing the raw material, those producing particular types of goods, and
the wholesale sellers. That is, this market supplies consumers in different
districts with practically the same set of goods.

The latter assumption allows us to treat the given market as a collection of
levels made up of firms with similar activities. Furthermore, we can specify
the density of each level of identical firms (for example level $i$) by $%
\rho _i({\bf r})$ and the material flow through one firm by $x_i({\bf r})$.
The levels are ordered according to the power of the firms and the higher
the level, the fewer the total number of firms at that level. Each firm buys
the product of firms at the level just above it and sells its own product to
firms in the next lower level. The highest level consists of the firm that
extracts the raw material, and the lowest one is made up of retail sellers.
Therefore, each level $i$ also contains micromarkets of products made by
those firms and, thus, should be characterized by a spatial distribution of
prices, $p_i({\bf r})$.

The conservation of materials at each level allows us to write 
\begin{equation}
\label{m1}x_i({\bf r})\rho _i({\bf r})=\int d{\bf r}^{\prime }G_{{\bf r},%
{\bf r}^{\prime }}^{i,i-1}x_{i-1}({\bf r}^{\prime })\rho _{i-1}({\bf r}%
^{\prime }). 
\end{equation}
Here $G_{{\bf r},{\bf r}^{\prime }}^{i,i-1}$ is the function specifying the
trade interaction between firms at levels $i$ and $i-1$ and is localized in
the domain controlled by the individual firms at level $i$. In particular, 
$$
\int d{\bf r}^{\prime }G_{{\bf r},{\bf r}^{\prime }}^{i,i-1}=1. 
$$
It should be noted that expression (\ref{m1}) reflects the fact that the
higher the level, the larger the domain of control of firms below it. At the
lowest level (level 1, the retail sellers) the flow of goods obeys the
equality 
\begin{equation}
\label{m1a}x_1({\bf r})\rho _1({\bf r})=S(p_1({\bf r})\mid {\bf r}), 
\end{equation}
where $S(p_1({\bf r})\mid {\bf r})$ is a given function of the consumer
demands.

The activity of each firm results in the profit \cite{ec2} 
\begin{equation}
\label{m2}\pi _i({\bf r})=\left[ p_i({\bf r})-p_{i+1}({\bf r})\right] x_i(%
{\bf r})-t_i\left( {\bf r}\mid x_i({\bf r})\right) , 
\end{equation}
where the function $t_i({\bf r}\mid x_i)$ quantifies the total cost of the
production activity of firms at level $i$ that are localized in the region $%
{\bf r}$. For the highest level ($N$), $p_{N+1}({\bf r})=0$. The cost $t_i(%
{\bf r}\mid x)$ is a convex function of its argument, $x$, i.e. the curve $%
t_i({\bf r}\mid x_i)$is slopes upward,\ and 
\begin{equation}
\label{m2a}\frac{\partial t_i}{\partial x}>0,\frac{\partial ^2t_i}{\partial
x^2}>0. 
\end{equation}
The function also takes into account the fixed cost, that is 
\begin{equation}
\label{m2b}t_i({\bf r}\mid 0)>0. 
\end{equation}

The interaction of trade between different levels will be specified by an
equilibrium in the supply-demand relations such that each firm maximizes its
own profit, 
\begin{equation}
\label{m3}\frac{\partial \pi _i}{\partial x_i}=0, 
\end{equation}
and the market is assumed to be characterized by perfect competition, 
\begin{equation}
\label{m4}\pi _i=0. 
\end{equation}
The last equality implies that there is no barrier to any firm entering or
leaving the market.

Equations (\ref{m3}) and (\ref{m4}) constitute the essence of the proposed
model for self-regulation of such an hierarchically organized market. We now
show that, under the given assumptions, the price of any one type of goods
does not depend on the demand for other goods.

As follows from expressions (\ref{m2}), (\ref{m2a}), and (\ref{m2b}), there
is a unique solution of the system of equations (\ref{m3}) and (\ref{m4}): $%
x_i^{*}({\bf r})$, $\Delta p_i({\bf r})\stackrel{\text{def}}{=}\left[ p_i(%
{\bf r})-p_{i+1}({\bf r})\right] $ meeting the conditions\cite{e4b}

\begin{equation}
\label{m5a}\left. \frac{\partial \ln \left[ t_i({\bf r}\mid x)\right] }{%
\partial \ln x}\right| _{x=x_i^{*}({\bf r})}=1, 
\end{equation}
\begin{equation}
\label{m5b}\Delta p_i({\bf r})=\left. \frac{\partial t_i({\bf r}\mid x)}{%
\partial x}\right| _{x=x_i^{*}({\bf r})}. 
\end{equation}
The value $x_i^{*}({\bf r})$ and the corresponding value $\Delta p_i({\bf r}%
) $ depend solely on the properties of the function $t_i({\bf r}\mid x_i),$
which reflects the efficiency of production. Therefore, because firms at the
highest level extract the raw material rather than buy it ($p_{N+1}({\bf r}%
)=0$), all prices at each level in such a perfect market are specified by
the efficiencies of their technological processes and not on their demands.
The demand by ultimate consumers for goods at the lowest level determines
the total flow of products through the levels. It follows in this case from (%
\ref{m1}), (\ref{m1a}) that the demand alone determines the density of firms
at each level. Therefore, variations in the consumer demand for one type of
goods have no effect on the price and flow of goods of another type.

\section{Closing remarks}

We realize that the models for ecological and economic systems that we have
considered are quite simplistic and cannot be applied directly to real
systems. Rather, our goal here has been to elaborate the mechanism of
self-regulation, which, we believe, is inherent in every natural system.
Such a mechanism is required by all natural living systems because of their
complex organization and the necessity that at each level they adapt to
changes in the environment. Indeed, the very complex organization of
ecological or economic systems implies that none of their elements can
possess all the necessary information on how the system must adapt to
changes in the environment. Indeed, if each element were to interact with
every other one, it either would take an infinite time for the system to
adapt or the system as a whole would be unstable. One of the ways available
for such a system to avoid this problem is to organize itself in
hierarchical fashion. Unfortunately, such organization might also cause the
system to acquire undesirable nonlocal interactions that are mediated
through by higher levels. To suppress such interaction there must be some
cooperative mechanism for system self-regulation. In our opinion, this
self-regulation is implemented by the active behavior of elements at each
level. Each element acts according to only its own goal, responding to only
the small amount of information it receives. However, the law of material
conservation acting across the hierarchical organization leads to the
self-processing of information. Thus, the small amount of information
available to each element informs it in an aggregated and implicit way about
the state of the system as a whole. Through such a cooperative way the
individual behavior of different elements is made consistent across levels
and enables the system to respond properly to changes in the environment 
\cite{ec4},\cite{e4a}.

As concerns ecosystems, we hypothesize that this mechanism for
self-regulation arises from the preference by animals to move in the
direction of increasing prey density and to avoid regions with an increasing
number of predators. The latter response dampens variations in the species
population which otherwise could become critical, because the higher the
population of one prey, the greater the extent its predators will specialize
in hunting them to the exclusion of others. 

In economic systems, each firm attempts to maximize its own profit, so if 
the total profit increases in the neigbouring region (either of space or
type of goods), firms will tend to relocate (or retool) into this region.
Such active behavior gives rise to variations in the density of firms. We
have related this active behavior to the condition that the total profit be
zero, due to the presence of perfect competition.

\newpage

\begin{figure}[p]
\unitlength=1.00mm
\special{em:linewidth 0.4pt}
\linethickness{0.4pt}
\begin{picture}(128.00,93.00)
\emline{120.00}{30.11}{1}{70.00}{89.89}{2}
\emline{70.00}{89.89}{3}{20.00}{30.11}{4}
\put(30.00,30.00){\vector(0,1){12.15}}
\put(40.00,30.00){\vector(0,1){12.15}}
\put(50.00,30.00){\vector(0,1){12.15}}
\put(60.00,30.00){\vector(0,1){12.15}}
\put(70.00,30.00){\vector(0,1){12.15}}
\put(80.00,30.00){\vector(0,1){12.15}}
\put(90.00,30.00){\vector(0,1){12.15}}
\put(100.00,30.00){\vector(0,1){12.15}}
\put(110.00,30.00){\vector(0,1){12.15}}
\put(50.00,54.00){\vector(0,1){12.19}}
\put(95.00,42.00){\vector(0,1){12.19}}
\put(45.00,42.00){\vector(0,1){12.15}}
\put(55.00,42.00){\vector(0,1){12.15}}
\put(65.00,42.00){\vector(0,1){12.15}}
\put(75.00,42.00){\vector(0,1){12.15}}
\put(85.00,42.00){\vector(0,1){12.15}}
\put(60.00,54.00){\vector(0,1){12.15}}
\put(70.00,54.00){\vector(0,1){12.15}}
\put(80.00,54.00){\vector(0,1){12.15}}
\put(90.00,54.00){\vector(0,1){12.15}}
\put(65.00,66.00){\vector(0,1){12.15}}
\put(75.00,66.00){\vector(0,1){12.15}}
\put(70.00,78.00){\vector(0,1){11.03}}
\put(78.00,93.00){\makebox(0,0)[cc]{$3$}}
\put(100.00,75.00){\makebox(0,0)[cc]{$2$}}
\put(115.00,50.00){\makebox(0,0)[cc]{$1$}}
\put(128.00,35.00){\makebox(0,0)[cc]{$0$}}
\emline{70.00}{89.00}{5}{76.00}{92.04}{6}
\emline{80.00}{78.00}{7}{96.00}{76.13}{8}
\emline{97.00}{73.00}{9}{90.00}{65.81}{10}
\emline{100.00}{70.00}{11}{100.00}{54.19}{12}
\emline{110.00}{42.00}{13}{113.00}{46.88}{14}
\emline{120.00}{30.00}{15}{125.00}{33.12}{16}
\put(70.00,23.00){\makebox(0,0)[cc]{space region}}
\put(70.00,18.00){\vector(1,0){50.00}}
\put(70.00,18.00){\vector(-1,0){50.00}}
\put(20.00,15.00){\vector(0,1){15.11}}
\put(120.00,15.00){\vector(0,1){15.11}}
\put(39.00,9.00){\vector(0,1){6.91}}
\put(43.00,9.00){\vector(0,1){6.91}}
\put(47.00,9.00){\vector(0,1){6.91}}
\put(70.00,11.00){\makebox(0,0)[cc]{sun irradiation}}
\put(89.00,9.00){\vector(0,1){6.91}}
\put(93.00,9.00){\vector(0,1){6.91}}
\put(97.00,9.00){\vector(0,1){6.91}}
\put(70.00,89.00){\circle*{2.00}}
\special{em:linewidth 1.0pt}
\linethickness{1.0pt}
\emline{60.00}{78.00}{17}{80.00}{78.00}{18}
\emline{50.00}{66.00}{19}{90.00}{66.00}{20}
\emline{40.00}{54.00}{21}{100.00}{54.00}{22}
\emline{30.00}{42.00}{23}{110.00}{42.00}{24}
\emline{20.00}{30.00}{25}{120.00}{30.00}{26}
\end{picture}

\caption{ Trophic level representation, where: 0 -- phytoplankton, 1 --
small organisms, 2 -- organisms belonging to different classes, 3 -- large
predators. }
\label{fig1}
\end{figure}
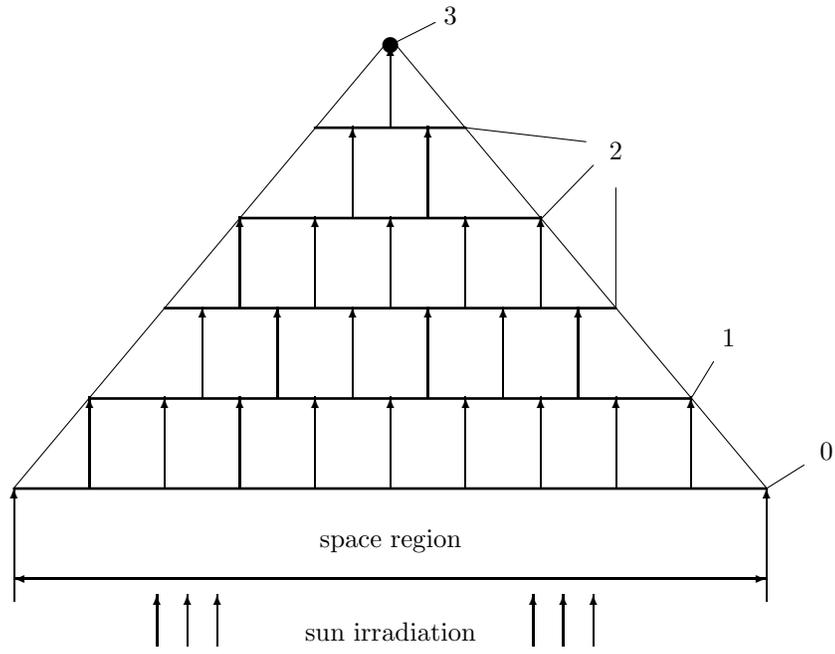

\end{document}